\documentclass[12pt]{iopart}
\usepackage{iopams}
\usepackage{graphicx}
\usepackage{amssymb}

\usepackage{graphicx}
\usepackage{dcolumn}
\usepackage{bm}
\usepackage[colorlinks=true,citecolor=blue,linkcolor=blue,urlcolor=blue]{hyperref}

\usepackage{tikz}
\usepackage{xcolor}
\usepackage{soul}
\usepackage{amsmath}

\newcommand{\orcidicon}[1]{\href{https://orcid.org/#1}{\includegraphics[height=\fontcharht\font`\B]{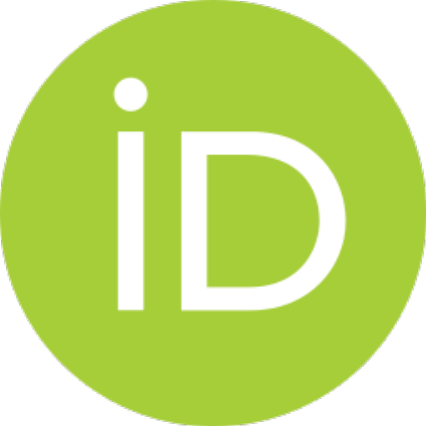}}}

\usepackage{pstricks}
\usepackage{color}
\usepackage{slashed}
\usepackage{epsfig}
\usepackage{amsfonts}
\usepackage[giveninits=true, sorting=none, backend=biber, style=numeric-comp, maxnames=10]{biblatex}
\bibliography{bib.bib}

\def\Re{\textrm{Re}}

\def\d{\mathrm{d}}
\def\q_perp{\mathbf{q}_{\perp}}
\begin{document}

\title[The generalized second law in Euclidean Schwarzschild black hole]{The generalized second law in Euclidean Schwarzschild black hole
}

\author{G. O. Heymans$^1$\footnote{On leave from Centro Brasileiro de Pesquisas Físicas - CBPF}, G, C. D. Rodríguez-Camargo$^2$, G. Scorza$^3$ and N. F. Svaiter$^3$}
\address{$^1$Institute of Cosmology, Department of Physics and Astronomy,
Tufts University, Medford, Massachusetts 02155, USA}

\address{$^2$Department of Physics and Astronomy, University College London, \\
London WC1E 6BT, United Kingdom}
\address{$^3$Centro Brasileiro de Pesquisas F\'{\i}sicas - CBPF, \\ 
Rua Dr. Xavier Sigaud 150, 22290-180 Rio de Janeiro, RJ, Brazil}

\eads{\mailto{olegario@cbpf.br},
\mailto{christian.rodriguez-camargo.21@ucl.ac.uk}
\mailto{gui.scorza@cbpf.br}, \mailto{nfuxsvai@cbpf.br}} 
\vspace{10pt}
\begin{indented}
\item[]October 2024
\end{indented}

\begin{abstract}
We discuss the Bekenstein generalized entropy of a Schwarzschild black hole, with the contribution of an external matter field affected by degrees of freedom near the event horizon. In the Euclidean section of the Schwarzschild manifold, we consider an Euclidean quantum effective model, a scalar theory in the presence of an additive disorder field. The average of the Gibbs free energy over the ensemble of possible configurations of the disorder is obtained by the distributional zeta-function method.  In the series representation for the average free energy, the effective actions give rise to generalized Schr\"{o}dinger operators on Riemannian manifolds. Finally, is presented the generalized entropy density with the contributions of the black hole geometric entropy and the external matter fields. The validity of the generalized second law using Euclidean functional methods is obtained.
\end{abstract}

%
%
%
%
%


\section{Introduction}\label{intro}

The aim of this work is to present a quite simple model with only a few assumptions to analyze the generalized second law in black hole thermodynamics \cite{PhysRevD.9.3292}. Many years ago, Bekenstein raised the following questions: What mechanisms ensure that the generalized entropy grows in any situation? And are there any exceptions to the law\cite{bekensteinn}? Since those questions were raised, a number of new developments have emerged \cite{davies1978thermodynamics,KAY199149,jacobson2003horizon,PhysRevD.81.064009,solodukhin2011entanglement,Carlip:2014pma,Hollands:2014eia,Harlow:2014yka,Raju:2020smc}. Using methods of constructive field theory and the distributional zeta-function method \cite{Svaiter:2016lha,svaiter2016disordered,polym12051066}, we are able to verify the validity of the generalized second law of thermodynamics in an unorthodox scenario.

The constructive program of quantum fields, also called Euclidean quantum field theory, uses Euclidean methods, with elliptic partial differential operators, to define Euclidean correlation functions \cite{symanzik1976modified,735b40cf-bd1b-33c4-9740-4aa21bbf9922,glimmquantum,jaffe1985euclidean}. This program is connected with the question: can quantum field theory be formulated using a classical probability theory \cite{kolmogorov1956foundations,kac1959probability,a61aa5fe-d74a-3133-bed2-f35c3c555015}?  
The program with probability measures has been implemented once that it is possible to perform an analytic continuation of a Lorentzian manifold to an Euclidean space with a positive defined metric \cite{PhysRev.75.1736,PhysRev.96.1124, schwinger1958euclidean,10.1143/PTP.21.241,symanzik1966euclidean}. After the aforementioned analytic continuation, the inhomogeneous Lorentz group becomes the Euclidean group, and commutative operators become random variables. Using the positive energy condition, the correlation functions of a scalar model, the so called Schwinger functions, are defined as the vacuum expectation values of products of the field operators analytically continued to the Euclidean region.
The Schwinger  functions have a classical probabilistic interpretation, that is, they are correlation functions between events in a probability space. In such approach, these correlation functions are moments of measures which are invariant under the Euclidean group.  For the case of a free scalar field, we have that the covariance operator is the Green's function of the  operator $(- \Delta + m_0^2)$ in ${\mathbb R}^{d}$, where $\Delta$ denotes the Laplacian and $m_0^2$ the spectral parameter. To go further we have to consider random linear functionals in ${\mathbb R}^{d}$ \cite{ito1954stationary, yaglom1957some, Gelfand:1959nq, Fradkin:1963vva}. The correlation functions in the Lorentzian spacetime are recovered using the Osterwalder--Schraeder axioms \cite{Osterwalder:1973dx,Osterwalder:1974tc}.

The limits of applicability of quantum field theory were put to the test by the formulation of quantum fields in curved spacetime, where problems of different nature appear  \cite{DEWITT1975295,Birrell_Davies_1982,fulling1989aspects, ford1997quantum}. After the concept of black hole entropy was introduced by Bekenstein \cite{Bekenstein:1972tm,PhysRevD.7.2333}, Hawking discussed free quantum fields in a fixed curved background spacetime geometry. It was proved that a black hole of mass ${M_{0}}$ emits thermal radiation at the temperature $\beta^{-1}$ which is proportional to the surface gravity of the horizon (a null hypersurface generated by a congruence of 
null geodesics) \cite{Hawking:1974rv, Hawking:1975vcx}.
This effect, originally derived  for a non-rotation neutral black hole is still a subject of a continuing debate 
and is a fertile ground to test new ideas and techniques.

The Euclidean methods discussed above play an important role in our understanding of black hole thermodynamics. The temperature of the black hole is proportional to the period of Euclidean time in the smooth Euclidean Schwarzschild manifold.
Here in this work we focus our attention in the following question: 
How can one use Euclidean quantum field theory to prove the generalized second law in the Euclidean section of Schwarzschild manifold? To answer such a question, one should first construct functional integrals in Riemannian manifolds. Such formulation is possible in ultrastatic spacetimes \cite{Wald:1979kp,Haag:1984xa,Jaffe:2006uz}. The main idea is to include the \textit{effects} of unknown degrees of freedom living 
near the event horizon over the external matter and radiation fields making use of disorder fields.

The microscopic degrees of freedom that are thought to contribute additional terms to a complete theory of black hole entropy have yet to be adequately identified \cite{page2005hawking}. Significant efforts have been made to explain the origin and behavior of these unknown contributions to entropy. Several common perspectives exist, including the statistical origin of Einstein's equations and the quantum properties of the gravitational field. For example, in Ref. \cite{chirco2014spacetime}, it is argued that an accurate interpretation of entropy does not necessarily require additional degrees of freedom but arises solely from the quantum nature of gravity.
Moreover, information theory has increasingly been invoked to support these discussions. Recent advances in higher-order networks, tied to information theory \cite{bianconi2021higher, Nokkala_2024}, provide a framework to apply these insights to quantum gravity on discrete geometries, such as simplicial and cell complexes. This allows for the extraction of modified gravity from generalized versions of entropy \cite{Bianconi_2021b, Bianconi_2021, Bianconi_2023, Bianconi_2024, bianconi2024gravityentropy}. Another alternative approach treats gravity as a purely classical background coupled to quantum fields \cite{oppenheim2022constraints, PhysRevX.13.041040, oppenheim2023gravitationally}. 
Finally, there are approaches that include the degrees of freedom of the black hole's interior \cite{maldacena}. These results, using the topological structure of replica wormholes, demonstrate why the black hole interior should be included in the computation of radiation entropy \cite{maldacena2013cool, gao2017traversable, almheiri2020replica, penington2022replica}.

We start recalling the claim done in \cite{PhysRevX.13.041040} where it is stated that the proposed randomness nature can be viewed as fundamental or as an effective theory of a deeper quantum gravity scheme, for instance, the structure given by the replica wormholes. In this paper, we present an approach which could be understood as a link between the aforementioned frameworks. In particular, we do construct a scheme which can be interpreted as a connection between the randomness nature of the degrees of freedom, and how this random behavior could model the \textit{effects} of the replica wormholes over the matter and radiation fields. Based in the Refs. \cite{Bombelli:1986rw,Callan:1994py,Kabat:1995eq,Terashima:1999vw,Das:2007mj}, we define in the Euclidean geometry an effective model with disorder. We are inspired in statistical field theory, where disorder has also been used for modelling systems with complex or unknown interactions \cite{PhysRev.107.333}. In realistic systems with random heterogeneties, quenched disorder refers to the degrees of freedom in which the relaxation time is much longer than the relaxation time for the system. Also, the probability distribution that defines quenched disorder is not affected by the degrees of freedom of the system \cite{ma2018modern}. In this case, the quenched free energy must be defined for systems with quenched disorder \cite{Brout:1959zz,klein1963statistical}. 

To take into account the influence of internal degrees of freedom, coming from the region near the event horizon to the generalized entropy, over the matter fields, we study a self-interaction $\lambda\varphi^{4}_{d}$ theory defined in an Euclidean section of the Schwarzschild  manifold.
In the Euclidean theory defined in a compact domain, we introduce an additive disorder field to model, in average, the \textit{effects} of the ``replica wormholes" over the matter and radiation fields. Once that we are modeling the \textit{effects} of quantities that  originate inside the event horizon, we suppose that the probability distribution does not depend on the distribution of the matter or radiation field. This leads us to using a quenched disorder to model these \textit{effects} on the matter fields.

There are numerous methods in the literature for calculating the quenched average of the Gibbs free energy. In this work, we utilize the distributional zeta-function method \cite{Svaiter:2016lha}. This method leads to a series representation of the quenched Gibbs free energy, allowing the imposition of various symmetries on the field theory encoded in each term of the series. In our procedure, see Sec. \ref{sec:thermal mass}, we are able to identify the variables which take into account the effects of the ``replica wormholes". Following this path, for specific choices of the disorder covariance, the problem is elevated to the domain of spectral theory involving singular differential operators on Riemannian manifolds. By applying this procedure, we find that generalized Schrödinger operators on Riemannian manifolds naturally emerge in the description of field thermodynamics quantities. In the context of a disorder field, the self-adjointness of the Schrödinger operator, defined by the effective actions, must be addressed \cite{shubin1992spectral,zbMATH02105661}. To the best of our knowledge, this connection is novel in the literature. Given that we are working in a compact domain, the Schrödinger operators have discrete spectra. With these countable sets of eigenvalues, we are able to define a spectral entropy.

Finally, using functional determinants, we derive a generalized entropy density, and our main result is a new expression for the generalized total entropy of the system affected by the additive disorder which, in average, could model the ``replica wormholes" effects. Remarkably, in this model, the generalized entropy preserves the second law of thermodynamics in black hole physics. Moreover, generalizing this result to finitely or countably many fields is straightforward. It is worth noting that other approaches exist for proving the generalized second law, as discussed in Refs. \cite{Sorkin:1986mg, Wall:2011hj,Wall:2009wm}.

The structure of this paper is as follows. In Sec. \ref{sec:disoderedLG} we discuss a self-interacting scalar field in Euclidean section of the Schwarzschild manifold in the presence of a disorder field. In section \ref{sec:thermal mass} we implement  
 the distributional zeta-function to computing 
the average of the generating functional of connected correlation functions. From our choice of the covariance of the disorder, naturally arises the theory of generalized Schr\"{o}dinger operators in Riemannian manifolds. 
In section \ref{sec:thermal mass2} the generalized entropy density of the black hole is discussed and the validity of the second law of thermodynamics is proved. Conclusions are given in Sec. \ref{sec:conclusions}. 
We use the units $\hbar=c=k_{B}=1$.

\section{The $\varphi_{d}^{4}$ Euclidean Field theory in a  random environment}\label{sec:disoderedLG}

The Birkhoff theorem on manifolds ensures that 
any vacuum spherical symmetric solution of the Einstein equation is locally isometric to a region in Schwarzschild spacetime. Therefore we start from the pseudo-Riemannian manifold with the  Schwarzschild metric in a $d$-dimensional spacetime \cite{Myers:1986un}. The line element reads

\begin{eqnarray}
   \d s^2 =& -\left(1 - \left(\frac{r_s}{r}\right)^{d-3}\right) \d t^2 + \left(1 - \left(\frac{r_s}{r}\right)^{d-3}\right)^{-1} \d r^2 + r^2\d \Omega^2_{d-2}. 
\end{eqnarray}
The Schawrzschild radius $r_s$ is proportional to the product of the $d$-dimensional Newton's constant and the black hole mass $M_{0}$,
\begin{eqnarray}
r^{d-3}_s = \frac{8\Gamma(\frac{d-1}{2})}{(d-2)\pi^{\frac{(d-3)}{2}}}G^{(d)} M_{0}.
\end{eqnarray}
For simplicity, in the following we use $G^{(d)}M_{0}=M$. With such a definition, in four dimensions, the quantity $M$ has units of length.

After a Wick rotation, $t \to i\tau$, in the time coordinate we obtain the $d$-dimensional Hawking instanton, i.e., a positive definite Euclidean metric for $r>r_s$. 
 \begin{eqnarray}
\d s_{E}^2 &= \left(1 - \left(\frac{r_s}{r}\right)^{d-3}\right) \d \tau^2 + \left(1 - \left(\frac{r_s}{r}\right)^{d-3}\right)^{-1} \d r^2 + r^2\d \Omega^2_{d-2}.
\end{eqnarray}
This manifold has a conic singularity. The singularity in $r=r_s$ is removed if we assume that the imaginary time coordinate, $\tau$, is a periodic coordinate with period $4\pi r_s/(d-3)$. 
The bifurcate Killing horizon becomes a rotation axis. This Euclidean section of the Schwarzschild solution, with compactified imaginary time, is homeomorphic to
$\mathbb{R}^{2}\times S^{2}$.

In such a manifold one defines the Israel-Hawking-Hartle vacuum state. Any quantum field defined in this manifold behave as if they are being held at a temperature $\beta^{-1}=(d-3)/4\pi r_s$. In the Matsubara formalism, the periodicity in imaginary time is associated to finite temperature states, where the Euclidean space is homeomorphic to $S^{1}\times \mathbb{R}^{3}$ \cite{kadanoff1989,Landsman:1986uw}. Since that, at principle, we do not have mathematical control of our expressions on the infinite volume limit. One need to enclose the black hole within a finite-volume box imposing some boundary conditions. From now on, we assume Dirichlet boundary conditions on the surface of the confining box. The volume total of the system is Vol$_{d}(\Omega) =\beta\,V_{d-1}$.
Here we would like to point out that in the case of Euclidean interacting field theories confined in compact domains it is necessary to introduce surface counterterms to make interacting field theories pertubatively renormalizable \cite{Symanzik:1981wd,Diehl:1981zz,Fosco:1999rs,Caicedo:2002ft, Svaiter:2004ad, AparicioAlcalde:2005wxe}.

In the following we do need to define our operators in the Riemannian manifold. Once in what follows we are considering a scalar field, we shall need to define the Lapace-Beltrami operator. In any smooth connected $d-$dimensional Riemannian manifold, $\mathcal{M}^d$, such an operator is defined by
\begin{equation}
-\Delta_{g}=-\frac{1}{{\sqrt g}}\sum_{i,j=1}^{d}\frac{\partial}{\partial x^{i}}\left({\sqrt g}g^{ij}\frac{\partial}{\partial x^{j}}\right),
\end{equation}
where $(g^{ij})=(g_{ij})^{-1}$, and $g= \mathrm{det}(g_{ij})$. We are working in a local arbitrary curvilinear coordinate system $x_{\nu}=(x_{1}, x_{2},...,x_{d})$. As usual, let us define the Riemannian $d$-volume $\mu$ defined by $\d \mu={\sqrt g}\,\d x_{1}\d x_{2}...\d x_{d}$. In general, we are interested in the Hilbert space of square integrable functions defined on a compact domain, that is, $\mathcal{H} = L^2(\Omega, d\mu)$, where $\Omega \subseteq \mathcal{M}^d$ is compact.

Using the fact that in the case of an interacting field theory, the black hole can remain in thermal equilibrium with a thermal bath \cite{Gibbons:1976es}, here we
consider a Euclidean self-interacting scalar model.  The action functional for a single self-interacting scalar field is given by  
\begin{equation}\label{eq:scalar}
S(\varphi)=\frac{1}{2}\int_{\beta} d\mu
\left[\varphi(x)\left(-\Delta_{s}+m_{0}^{2}\right)\varphi(x)+\frac{\lambda_{0}}{12}\varphi^{4}(x) \right].
\end{equation}
The symbol $-\Delta_{s}$ denotes the Laplace-Beltrami operator in the Euclidean section of the Schwarzschild manifold ${\cal{M}}^{d}_{s}$, $\lambda_{0}$ the bare coupling constant and $m_{0}^{2}$ spectral parameter of the model. Also, the notation $\int_{\beta}$ means that we have a periodic imaginary time coordinate $x_{1}=\tau$, that is,  $0\leq x_{1}\leq 4\pi r_s/(d-3)$. Therefore, $\varphi(x_{1},x_{2},x_{3},...,x_{d}) = \varphi(x_{1}+\beta,x_{2},x_{3},...,x_{d})$. We define $x_{2}=r$, as the radial coordinate. In such a manifold, the Laplace-Beltrami operator is explicitly given by
\begin{align}\label{eq:lbs}
 -\Delta_{s}\varphi =\Delta_{\theta}\varphi(x_{3},...,x_{d})&+\left(1-\left(\frac{r_{s}}{x_2}\right)^{d-3}\right)^{-1}\frac{\partial^{2}\varphi}{\partial x_{1}^{2}} \nonumber \\
 &+\frac{1}{x_2^{d-2}}\frac{\partial}{\partial x_2}\left(x_2^{d-2}\left(1-\left(\frac{r_{s}}{x_2}\right)^{d-3}\right)\frac{\partial\varphi}{\partial x_2}\right). 
\end{align}
where $\Delta_{\theta}$ denotes the Laplace-Beltrami operator in 
the $S^{d-2}$ the $(d-2)$-dimensional unit sphere, i.e., the contribution from the angular part. Finally, as previously stated, we are assuming  Dirichlet boundary conditions. We write $\varphi(x)|_{\partial{\cal M}^{d}_{s}} = 0$, since we are considering the whole system inside a reflecting wall. Such a procedure is needed by virtue of the fact that the system must have a finite volume and the spatially cut-off Schwinger function must be well defined

By introducing an external source $j(x)$, one can define the generating functional of all $n$-point correlation functions $Z(j)$ as \cite{rivers1988path,zinn2021quantum}
\begin{equation}
Z(j)=\int [\d \varphi]\,\, \exp\left(-S(\varphi)+\int_{\beta}
\d \mu\,j(x)\varphi(x)\right),
\label{eq:generatingfunctional}
\end{equation}
where $[\d \varphi]$ is a functional measure given by $[\d \varphi]=\prod_{x\, \in \, S^1 \times \mathbb{R}^3} \d \varphi(x)$. It is clear that $[\d \varphi]$ has only a symbolic meaning.
The next step is to define the generating functional of connected correlation functions $W(j)=\ln Z(j)$ and the Gibbs free energy $\ln Z(j)|_{j=0}$. Now we are able to construct our model. To take into account the \textit{effects} of the ``replica wormholes" over the scalar field entropy we introduce a quenched disorder field in which we are taking averages. The link between wormholes and quenched disorder is not new in the literature. In a different context, Ref. \cite{Heymans:2023tgi} proposed an anisotropic disorder to construct an analog model to Euclidean wormholes \cite{Hawking:1988ae, Coleman:1988tj, Klebanov:1988eh}.
The action functional for the scalar field in the presence of the disorder is given by
\begin{equation} 
S(\varphi,h)=S(\varphi)+ \int_{\beta} \d \mu\,h(x)\varphi(x),
\end{equation}
\noindent where  $S(\varphi)$ is the Euclidean action functional defined in Eq. (\ref{eq:scalar}), and $h(x)$ is the additive quenched disorder. 
At this point, let us introduce the functional $Z(j,h)$, the generating functional of correlation functions in the presence of disorder. Here we use an auxiliary external source $j(x)$, to generate the $n$-point correlation functions of the model. We define $Z(j,h)$ as follows
\begin{equation}
Z(j,h)=\int[\d\varphi]\, \exp\left(-S(\varphi,h)+\int_{\beta} \d \mu\,j(x)\varphi(x)\right).
\label{eq:disorderedgeneratingfunctional}
\end{equation}

As in the case without disorder, one can define a generating functional of connected correlation functions in the presence of disorder, i.e., the generating functional of connected correlation functions for one disorder realization, $W(j,h)=\ln Z(j,h)$. For the case of the disorder field, one can define an average generating functional of connected correlation functions, performing the quenched average over the ensemble of all realizations of the disorder. It is written as
\begin{equation}
\mathbb{E}\bigl[W(j,h)\bigr]=\int\,[\d h]P(h)\ln Z(j,h),
\label{eq:disorderedfreeenergy}
\end{equation} 
where $[\d h]$ is a functional measure, given by $[\d h]=\prod_{x} \d h(x)$,
and the probability distribution of the disorder field is written as $[\d h]\,P(h)$. This procedure is similar to the one used in statistical field theory where the free energy must be self-averaging over all the realizations of the disorder. In the next section, we shall discuss the method we employ to take such an average: the distributional zeta-function method. Also discussed are some conditions that a Schr\"odinger operator must satisfy to be well defined.

\section{The distributional zeta-function method}\label{sec:thermal mass}

A technical problem is how to calculate Eq. \eref{eq:disorderedfreeenergy}. There are many ways in the literature to perform such an average. Some of them are: the replica trick \cite{Edwards:1975aas,Emery:1975zz,mezard1987spin,dotsenko2005introduction}, the dynamical \cite{PhysRevB.18.4913,PhysRevLett.47.359,de2006random}, and the supersymmetric approaches \cite{Efetov:1983xg}. An alternative method 
that has been discussed in the literature is the distributional zeta-function method.

For a general disorder probability distribution, using the functional integral $Z(j,h)$ given by Eq. \eref{eq:disorderedgeneratingfunctional}, the distributional zeta-function, $\Phi(s)$, is defined as
\begin{equation}\label{eq:pro1}
\Phi(s)=\int [\d h]P(h)\frac{1}{Z(j,h)^{s}},
\end{equation}
for $s\in \mathbb{C}$, this function is defined in the region where the above integral converges. The average generating functional can be written as 
\begin{equation}
\mathbb{E}\left[W(j,h)\right]=-\left.\frac{\d}{\d s}\Phi(s)\right|_{s=0^{+}}, \qquad \Re(s) \geq 0,  
\end{equation}
where it is defined the complex exponential as $n^{-s}=\exp(-s\log n)$, with $\log n\in\mathbb{R}$.
Using analytic tools, the average generating functional of connected correlation functions, or the quenched free energy, can be represented as
\begin{eqnarray}\label{eq: m23e}
\mathbb{E}\bigl[W(j,h)\bigr]&=\sum_{k=1}^{\infty} \frac{(-1)^{k+1}a^{k}}{k k!}\,
\mathbb{E}\,[(Z(j,h))^{\,k}]-\ln(a)+\gamma+R(a,j),
\end{eqnarray}
\noindent where $a$ is a dimensionless arbitrary constant, $\gamma$ is the Euler-Mascheroni constant, and $|R(a)|$ is given by
\begin{equation}
R(a,j)=-\int [\d h]P(h)\int_{a}^{\infty}\,\frac{\d t}{t}\, \exp\left(-Z(j,h)t\right).
\end{equation}
%
For large $a$, the quantity $|R(a)|$ is quite small, therefore, the dominant contribution to the average generating functional of connected correlation functions is given by the moments of the generating functional of correlation functions of the model. In previous works, it has been shown that the limit of large $a$ is associated to the thermodynamic limit \cite{Rodriguez-Camargo:2022wyz, Heymans:2024dzq}. Later, we will absorb such a quantity in the total volume. Hence, assuming that $a$ is large enough, we can write \eref{eq: m23e} as
\begin{equation}\label{eq:feq}
    \mathbb{E}\left[W(j,h)\right]=\sum_{k=1}^{\infty} c_k \, \mathbb{E}\left[Z^k(j,h)\right],
\end{equation}
where we have defined $c_k = \frac{(-1)^{k+1}a^{k}}{k k!}$.

 Once that the average is taken, the covariance of the disorder field must be chosen with some care. If one chooses a Gaussian disorder, all the points of the Euclidean manifold will feel the effects in the same way. However, it does not seem to cover our interest to justify the black hole as the physical source of such disorder. Since we are in the simplest case of an Euclidean Schwarzschild black hole, we must expect that the effects increase in the vicinity of the black hole horizon. For this reason, we choose the the covariance of the disorder to be given by 
\begin{equation}
\mathbb{E}[{h(x)h(y)}]=\frac{V(x_{2})}{\sqrt g}\delta^{d}(x-y),
\end{equation}
where we are assuming that the functional form of $V(x_{2})$ is
\begin{equation}\label{vofx}
    V(x_{2})=b^{\alpha-2}(x_{2})^{-\alpha}
\end{equation}
for $\alpha$ positive definite, and $b$ being a constant with units of length. We do remark that $x_{2}=r$, therefore Eq. \eref{vofx} has a spherical symmetry. After integrating over all the realizations of the disorder we get that each moment of the generating functional of connected correlation functions $\mathbb{E}\,[Z^{k}(j,h)]$ can be written as
\begin{equation}
\mathbb{E}[Z^{k}(j,h)]=\int\prod_{i=1}^{k}[\d \varphi_{i}^{(k)}]\,\exp\left(-S_{\textrm{eff}}(\varphi_{i}^{(k)},j_{i}^{(k)})\right),
\label{aa11}
\end{equation}
\noindent where the effective action $S_{\textrm{eff}}\left(\varphi_{i}^{(k)}\right)$ is describing the field theory with $k$-field components. From now on, we are omitting the super-index $k$ within the fields. In this new notation the effective action is given by
\begin{align}\label{eq:Seff}
S_{\textrm{eff}}^{(k)}\left(\varphi_{i},j_{i}\right) =
\int_{\beta} \d \mu&\left[\sum_{i=1} ^k\left(\frac{1}{2}\varphi_{i}(x)\left(-\Delta_{s}+m_{0}^{2}\right)\varphi_{i}(x)
+\frac{\lambda_{0}}{4!}\left(\varphi_{i}(x)\right)^{4}\right)\right.\nonumber\\
&\left.-\frac{V(x_{2})}{2}\sum_{i,j=1}^k\varphi_{i}(x)\varphi_{j}(x)
-\sum_{i=1}^k\varphi_{i}(x)j_{i}(x)\right].
\end{align}

A remarkable aspect of this formalism is that after the averaging procedure with a reduced description of these degrees of freedom, new collective variables are appearing, i.e., multiplets of fields in all moments. In some previous works, it was used the following configuration of the scalar fields: $\varphi_{i}(x)=\varphi_{j}(x)$ in the function space, and also $j_{i}(x)=j_{j}(x)$ $\forall \,i,\,j \in \mathbb{R}^k$.  All the terms of the series have the same structure and it is minimized each term of the series one by one \cite{Diaz:2017grg,Diaz:2017ilf,Rodriguez-Camargo:2021ryf,Rodriguez-Camargo:2022wyz,Heymans:2022sdr}. 

It is worth noticing that the term which can be interpreted as the ``replica wormholes" is given by the contribution proportional to $\varphi_i\varphi_j$. Thus, the aforementioned ansatz is not able to capture all the contributions to the physical quantities, leading to the need of an ansatz free approach. In different context, the ansatz free approach has been constructed and employed to describe various kind of physical situations \cite{Heymans:2023tgi, Heymans:2024dzq,antonio}. Since we are interested only in the thermodynamic properties of the model, we have no need to generate the correlation functions. Therefore, we set $j_i(x)=0$, $\forall$ $i$ and we omit the $j=0$ argument in all quantities. 

Let us discuss the Gaussian contribution of the action given by Eq. (\ref{eq:Seff}), once that it is enough to access the thermodynamic properties. The free part of the effective action can be recast as
\begin{align}
S_{0}^{(k)}\left(\varphi_{i}\right)= \frac{1}{2}\int_{\beta} \d \mu \sum_{i,j=1}^{k}\varphi_{i}(x)\left[\left(-\Delta_{s}+m_{0}^{2}\right)\delta_{ij}-V(x_{2})\right]\varphi_{j}(x).
\end{align}
The  differential operator is not diagonal in the $(i,j)$-space. In order to diagonalize it, let us construct the following $k\times k$ matrix: 
\begin{equation}\label{matrixg1}
    G \equiv \left[G_{ij}\right] \equiv \left[\begin{array}{cccc}
        G_{11} - V & -V & \cdots & -V  \\
   -V & G_{22} -V & \cdots & -V  \\
   \vdots & \cdots & \ddots & \vdots \\
   -V & -V & \cdots & G_{kk} - V  \\
  \end{array}\right] .
\end{equation}

In the entries of the matrix \eref{matrixg1} we have used the following definition
\begin{equation}
    G_{ij} \equiv \left(-\Delta_{s} + m_0^2\right)\delta_{ij}.
\end{equation}
The matrix $G$ is a symmetric matrix since $V(x_{2})$ is a real-valued function. Then $G$ can be diagonalized by an orthogonal transformation, $S$. Defining $A = \langle S , G S \rangle$ as the diagonal matrix, it can be shown that the matrix $A$ is given by
\begin{equation}\label{eq:diagprop}
    A = \left[
\begin{array}{cccc}
  G_{11} -kV & 0& \cdots & 0 \\
   0 & G_{22} & \cdots & 0 \\
   \vdots & \cdots & \ddots & \vdots \\
   0 & \cdots & & G_{kk}
\end{array}
\right].
\end{equation}
Defining $\Phi$ as the vector of components $\varphi_i$, we can use the matrix $S$ to construct the vector $\tilde{\Phi} = S\Phi$. Denote the components of the vector $\tilde{\Phi}$ by $\phi_i$. It can be checked directly that each $\phi_i$ is given by a linear combination of $\varphi_j$, with $j\leq i$. Such an observation allow us to identify the variables $\phi_i$ as the variables which take into account the effects of the ``replica wormholes". Disregarding the interacting action, one can use the fact that the the matrix $S$ is orthogonal and the form of $A$ given by Eq. (\ref{eq:diagprop}) to write

\begin{eqnarray}
\mathbb{E}\left[Z^{k}(h)\right]&=\int[\d \phi]\,\exp\left(-S^{(k)}(\phi)\right)\int\,\prod_{l=2}^{k}\left[\d \phi_{l}\right]\,\exp\left(-S_{0}^{(k)}(\phi_{l})\right),
\end{eqnarray}
where we are denoting $\phi_1 = \phi$, 
\begin{equation}\label{eq:SV}
S^{(k)}(\phi)= \frac{1}{2}\int_{\beta} \d \mu \,\phi(x)\left[-\Delta_{s}+m_{0}^{2}-kV(x_2)\right]\phi(x),
\end{equation}
and
\begin{equation}\label{eq:S0}
S_{0}^{(k)}(\phi_{l})= \frac{1}{2}\int_{\beta} \d\mu \sum_{l=2}^{k} \phi_{l}(x)\left(-\Delta_{s}+m_{0}^{2}\right)\phi_{l}(x).
\end{equation}

Performing all the Gaussian integrations we can rescast our quenched Gibbs free energy, Eq. (\ref{eq:feq}), in an enlightening form
\begin{equation}\label{eq:finalfeq}
    \mathbb{E}\left[W(h)\right] = \sum_{k=1}^{\infty}c_k \, \left[\textrm{det}\left(-\Delta_s + m_0^2\right)\right]^{\frac{1-k}{2}} \left[\textrm{det}\left(-\Delta_s + m_0^2 - kV(x_2)\right)\right]^{-\frac{1}{2}}.
\end{equation}
Notice that the first determinant is a usual one. Its structure is expected in the analyses of scalar fields on a Riemannian manifold. The regularity and self-adjointness of such an operator follows from the regularity of the Laplace-Beltrami operator. However, the second determinant describes a more complex situation. Such an operator is a Schr\"{o}dinger operator on a Riemannian manifold. 

The self-adjointness of the Schr\"{o}dinger operator on a Riemannian manifold
in a Hilbert space alongside with its spectral properties must be determined. For the $(-\Delta)$ 
in $L^{2}(\mathbb{R}^{d})$, the Fourier transform establishes self-adjointness on a domain ${\cal{D}}(-\Delta)=H^{2}(\mathbb{R}^{d})$, which corresponds to the Sobolev space. If the 
Schr\"{o}dinger operator is not proven to be essentially self-adjoint, there may be an infinite set of self-adjoint extensions, and find the physical/correct one is a hard task \cite{zbMATH03389535,bams/1183538232, https://doi.org/10.1112/S0024609302001248,Diyab_Reddy_2022}.

Which condition must the potential satisfy to ensure that the Schr\"odinger operator is essentially self-adjoint? An important result was obtained by Oleikin \cite{zbMATH00687507}. This author proved that in the absence of local singularities in the potential, the  Schr\"{o}dinger operator in a Riemannian manifold is essential self-adjoint. Note that $V(x_{2})$ is a real-valued function which is locally summable in $L^{2}$ and globally semi-bounded, i.e., $V(x_{2})\geq -C$ for $x_{2} \in {\cal{M}}_{s}^{d}$, with a constant $C \in \mathbb{R}$. Therefore we have a self-adjoint operator in the Hilbert space $L^{2}({\cal{M}}^{d}_{s})=L^{2}({\cal{M}}_{s}^{d},d\mu)$. 
Using the result of this section, Eq. (\ref{eq:finalfeq}), we are able to express the generalized entropy density as the ratio between two functional determinants.

\section{The generalized entropy density}\label{sec:thermal mass2}

To preserve the universality of the second law of thermodynamics \cite{Bardeen:1973gs}, Bekenstein conjectured that the total entropy of the system must satisfy the generalized second law
\begin{equation}\label{eq:genentro}
\Delta S_{BH}=\Delta S^{(1)}+ \Delta S^{(2)}\geq 0,
\end{equation}
where $S^{(1)}$ is the Bekenstein-Hawking entropy, proportional to the horizon area and $S^{(2)}$, is the correction  
of matter and radiation fields.
We now proceed to discuss the contribution given by $S^{(2)}$.


Since, in our case, we have a system with infinitely many degrees of freedom, we must use the concept of mean entropy, i.e., the entropy per unit $(d-1)-$volume $(\beta^{-1}\textrm{Vol}_{d}(\Omega))$ \cite{zbMATH03233588},
\begin{equation}
s^{(2)}=\frac{\beta\,S^{(2)}}{\textrm{Vol}_{d}(\Omega)}. 
\end{equation}
Using the fact that $S=\ln Z+ \beta E$, in an Euclidean quantum field theory we can derive the generalized entropy density from the Gibbs free energy. In the case of a compact Riemannian manifold, the contribution of the quantum fields to the generalized entropy in the absence of the disorder is 
\begin{equation}
s^{(2)}=\frac{1}{\textrm{Vol}_{d}(\Omega)}\left.\left(\beta-\beta^{2}\frac{\partial}{\partial\beta}\right)\ln Z(j)\right|_{j=0},
\end{equation}
where $Z(j)|_{j=0}$ is the partition function. Here we have a Gibbs entropy of a classical probability distribution.
In the presence of disorder the contribution of external matter fields to the generalized entropy density $s^{(2)}$ is
\begin{equation}\label{s211}
s^{(2)}=\frac{1}{\textrm{Vol}_{d}(\Omega)}\biggl(\beta-\beta^{2}\frac{\partial}{\partial\beta}\biggr)\mathbb{E}\bigl[W(h)\bigr] .
\end{equation}

The form of Eq. \eref{s211} results from the assumption that the total volume and the temperature are not affected by the disorder. Using Eqs. (\ref{eq:finalfeq}) and (\ref{eq:feq}), we obtain that
\begin{align}\label{eq:entr}
s^{(2)}=\sum_{k=1}^{\infty}\frac{c'_{k}}{\textrm{Vol}_{d}
(\Omega)}
\left(\beta-\beta^{2}\frac{\beta}{\partial\beta}\right)
\left[\textrm{det}( -\Delta_{s} + m_{0}^2)\right]^{-\frac{k}{2}}\left[\frac{\textrm{det}\left(-\Delta_{s}+m_{0}^{2}\right)}{
\textrm{det}\left( -\Delta_{s} + m_{0}^2-kV(x_{2})\right)}\right]^{\frac{1}{2}},
\end{align}
where $c'_k = \frac{(-1)^k}{kk!}$. Notice that we have absorbed $a^k$ into the total volume.
%

The entropy, in physical grounds, depends on the covariance of the disorder. It becomes necessary to specify $V(x_{2})$ in virtue to obtain $s^{(2)}$. As we shall clarify  below, we will obtain the values of the functional determinants using their eigenfunctions. One can verify in Eq. (\ref{eq:lbs}), that the operator $\Delta_s$ contains always the angular Laplace-Beltrami, $-\Delta_{\theta}$. Since $V(x_2)$ does not depend on the angular variables, we shall ignore such an angular operator. In practice, it is equivalent to work in $d=2$. In the neighborhood of the event horizon is expected that the effects of the internal degrees of freedom become more relevant. Going to such a region, $x_{2}=r\approx 2M$, we can define the radial coordinate $\rho=\sqrt{8M(r-2M)}$, and the line element can be written as 
\begin{equation}\label{eq:RindlerA}
\d s^{2}=\frac{\rho^{2}}{16M^2}\d\tau^{2}+\d\rho^{2},
\end{equation}
where the horizon is located at $\rho=0$. The equation of motion for the free field in the Euclidean Rindler space is given by
\begin{eqnarray}
(-\Delta_{\textrm{R}} + m_0^2)\phi &=\left(\frac{16M^2}{\rho^{2}}\frac{\partial^{2}}{\partial\tau^{2}}+\frac{\partial^{2}}{\partial\rho^{2}}+\frac{1}{\rho}\frac{\partial}{\partial\rho}+m_{0}^{2}\right)\phi=0,
\end{eqnarray}
where $-\Delta_{\textrm{R}}$ stands for the Laplace-Beltrami operator in the Rindler coordinates given by the line element (\ref{eq:RindlerA}). Therefore, we can observe that this operator is $-\Delta_s$ near the horizon after the angular part is disregarded.

In the near-horizon approximation, that is $\rho \approx 0$, the potential of the Schr\"{o}dinger operator can be recast as

\begin{equation}
V(\rho, M)=\frac{a^{\alpha-2}}{(2M)^{\alpha}}\biggl(1-\frac{\alpha\rho^{2}}{16M^{2}}\biggr).
\end{equation}

Using the fact that the coordinate $\tau$ is periodic, the total entropy density will be a sum over all the Matsubara modes
\begin{equation}
s^{(2)}_{BGH}=\sum_{n=-\infty}^{\infty}
s^{(2)}(n).
\end{equation} 
Where $s^{(2)}(n)$ is given by Eq. (\ref{eq:entr}) in the near the horizon approximation with the angular part disregarded.

Note that for small $\rho$ and defining $f(\alpha,M)=\alpha/2^{4+\alpha}M^{2+\alpha}$, the determinant which contains the potential can be written as
\begin{equation}
\textrm {det}\left[ -\Delta_{R}+ka^{\alpha -2}\rho^{2}f(\alpha,M) + m_{0}^2-\frac{ka^{\alpha -2}}{(2M)^{\alpha}}
\right].
\end{equation}
We do define an effective mass for each effective action given by $m_{\textrm{eff}}^2(k,M)= m_{0}^2-ka^{\alpha -2}/(2M)^{\alpha}$.
To continue, let us discuss the solution of the differential equation for each Matsubara mode. We have that $R_{n}(\rho)$ satisfies
\begin{equation}
\left[\rho^{2}\frac{\d^{2}}{\d\rho^{2}}+\rho\frac{\d}{\d\rho}+m_{\textrm{eff}}^{2}\rho^{2}-n^{2}\right]R_{n}(\rho)=0.
\end{equation}
Defining $w=m_{\textrm{eff}}^2\rho^2$, the general solution of the above equation is written as
\begin{equation}
R_n(x) = AJ_n(w)+BY_n(w),
\end{equation}
where $J_n(w)$ is the Bessel function of the first kind and $Y_n(w)$ is the Bessel function of the second kind. Using the fact that the large $n$ Matsubara modes give the main contribution to the generalized entropy~\cite{solodukhin2011entanglement}, we can write an asymptotic expansion for $J_n(w)$ and $Y_n(w)$. Since $m_{\textrm{eff}}^2(k,M)$ can be negative for some $k$ we write $s^{(2)}(n)$ as 
\begin{equation}\label{s2nkcs}
s^{(2)}(n)=s_{k<k_{c}}^{(2)}(n)+s^{(2)}_{k\geq k_{c}}(n).
\end{equation}
Denoting by $\lfloor m \rfloor$ the largest integer less or equal to $m$, we define a critical $k$ given by $k_{c}=\lfloor\frac{(2M)^{\alpha}m_{0}^{2}}{a^{\alpha -2}}\rfloor$. Using that $\beta = 8 \pi M$, we have
\begin{align}
s^{(2)}_{k<k_{c}}(n)=8 \pi\left(M-M^2\frac{\partial}{\partial M}\right)
\sum_{k=1}^{k_{c}-1}
\frac{c'_{k}}{\textrm{Vol}_{d}(\Omega)}\left[\textrm{det}( -\Delta_{R} + m_{0}^2)\right]^{\frac{-k}{2}}\left[\frac{\textrm{det}\left(-\Delta_{R}+m_{0}^{2}\right)}{
\textrm{det}\left( -\Delta_{R} + m_{\textrm{eff}}^2\right)}\right]^{\frac{1}{2}},
\end{align}
and
\begin{align}
s^{(2)}_{k\geq k_{c}}(n)=8 \pi\left(M-M^2\frac{\partial}{\partial M}\right)
\sum_{k=k_{c}}^{\infty}
\frac{c'_{k}}{\textrm{Vol}_{d}(\Omega)}\left[\textrm{det}( -\Delta_{R} + m_{0}^2)\right]^{\frac{-k}{2}}\left[\frac{\textrm{det}\left(-\Delta_{R}+m_{0}^{2}\right)}{
\textrm{det}\left( -\Delta_{R} + {m'}_{\textrm{eff}}^{2}\right)}\right]^{\frac{1}{2}},
\end{align}
where ${m'}^2_{\textrm{eff}} = -2m^2_{eff}$ is the shifted effective mass.

The spectrum of the Schr\"{o}dinger operator is unknown. Therefore, here we use an alternative procedure to calculate the above expression. It can be shown that the derivative of the spectral zeta function can be expressed in terms of the eigenfunctions as follows
\begin{equation}\label{eq:Gel}
-\left.\frac{d}{ds}\zeta(s)\right|_{s=0}=\ln\left[\frac{ R(0)}{R(-\infty)}\right],
\end{equation}
where $R$ denotes the respective eigenfunctions. This is know as the Gel'fand-Yaglom method, which consists in manipulate the eigenfunctions instead of the eigenvalues. Using this procedure it is possible to evaluate the generalized entropy density. We can evidence that an eigenfunction which is repeating in both limits will cancel out. This justifies the fact that we have disregarded the angular Laplace-Beltrami in Eq. (\ref{eq:RindlerA}). Since the eigenfunctions of such an operator are going to be spherical harmonics, we have that they are $\rho$-independent. For  $\alpha =2$ we obtain the following expression for the first contribution of Eq. \eref{s2nkcs}
\begin{eqnarray}
s^{(2)}_{k<k_{c}}(n)= \sum_{k=1}^{k_{c}-1}\frac{c'_{k}}{\textrm{Vol}_{d}}\left[\frac{2\pi kn  }{Mm^2_{\textrm{eff}}}+ 8 \pi M\right]\left(\frac{ m_{0}}{m_{\textrm{eff}}}\right)^{n} .
\end{eqnarray}

A similar result is obtained for the second contribution of Eq. \eref{s2nkcs}
\begin{eqnarray}
s^{(2)}_{k \geq k_{c}}(n)=
\sum_{k=k_{c}}^{\infty}\frac{c'_{k}}{\textrm{Vol}_{d}}\left[\frac{2\pi kn }{M{m'}^2_{\textrm{eff}}}
+ 8 \pi M\right]\left(\frac{ m_{0}}{{m'}_{\textrm{eff}}}\right)^{n}.
\end{eqnarray}

The generalized second law was introduced to ensure that the total entropy of the system also increases ($\Delta S^{(1)}+\Delta S^{(2)}\geq 0$). Departing from the Eq. \eref{s2nkcs}, we have that $ S^{(2)}(n)=S^{(2)}_{k<k_{c}}(n) + S^{(2)}_{k \geq k_{c}}(n) $. Thus, the expressions for both the contributions of the entropy yield

\begin{eqnarray}\label{eq:entro1}
S^{(2)}_{k<k_{c}}(n)=
\sum_{k=1}^{k_{c}-1}c_{k}\left[\frac{kn}{4M^2m^2_{\textrm{eff}}}
+ 1\right]\left(\frac{ m_{0}}{m_{\textrm{eff}}}\right)^{n} ,
\end{eqnarray}
for for $k \leq k_{c}$, and 
\begin{eqnarray}\label{eq:entro2}
S^{(2)}_{k \geq k_{c}}(n)=
\sum_{k=k_{c}}^{\infty}c_{k}\left[\frac{kn}{4M^2m'^2_{\textrm{eff}}}
+ 1\right]\left(\frac{ m_{0}}{m'_{\textrm{eff}}}\right)^{n},
\end{eqnarray}
for $k \geq k_{c}$. If we consider the two angular variables which have been disregarded the result is preserved, as can be observed by Eq. (\ref{eq:Gel}). Further corrections must be analyzed.

\begin{figure}[ht]
        \begin{center}
            \includegraphics[width=0.7\textwidth]{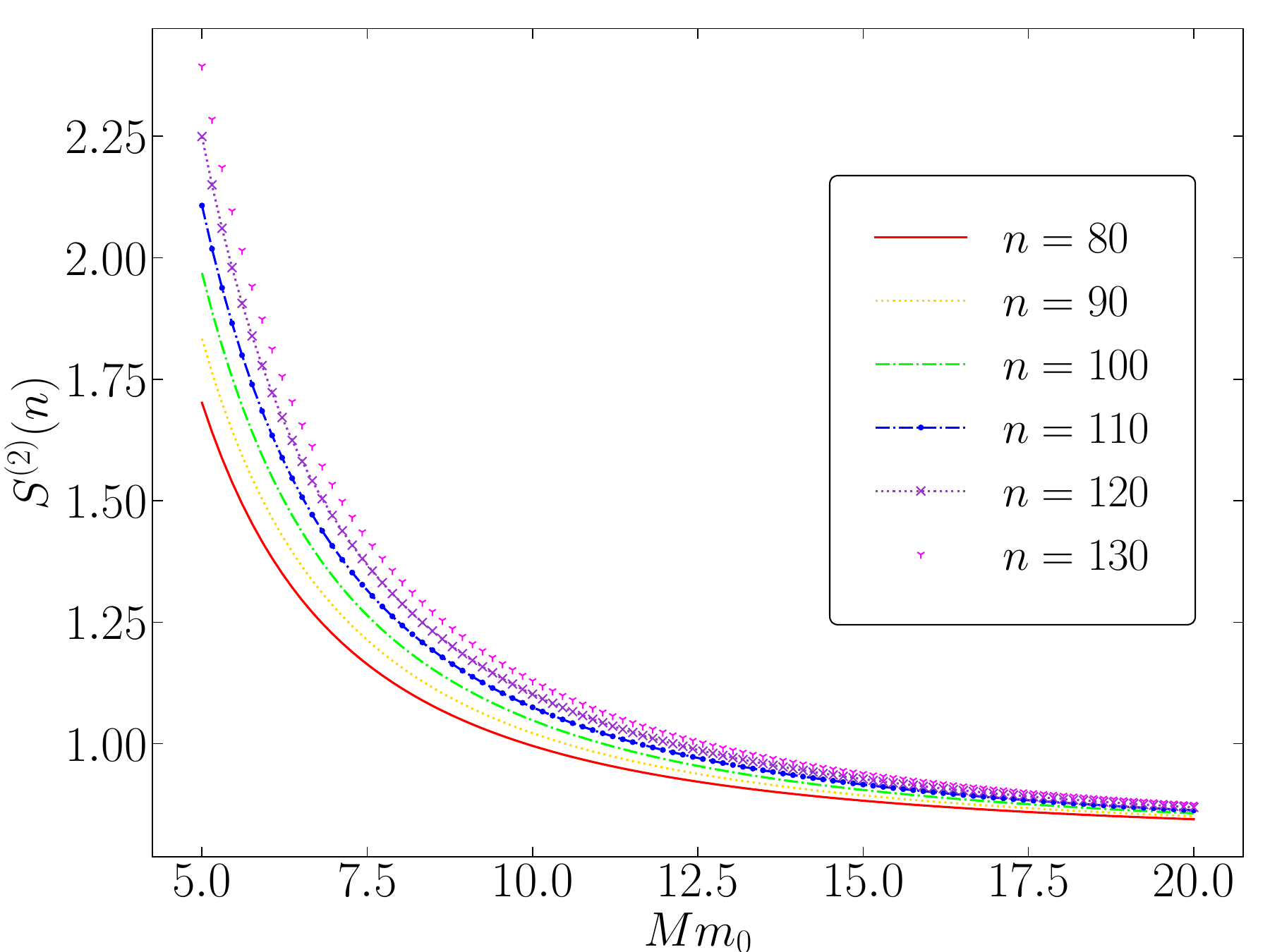}\\
        \end{center}
	\caption{Behaviour of the matter entropy as a function of the dimensionless parameter $Mm_{0}$ for different Matsubara modes $n$. We do remark the redefinition of $M$ as $M=G^{(d)}M_{0}$.}
	\label{figen1}
 \end{figure}

In order to figure out the numerical validity of our results, we evaluate our expressions for different situations. In Fig.~\ref{figen1} we plot the contributions for the sum of Eq. (\ref{eq:entro1}) and Eq. (\ref{eq:entro2}) for large Matsubara modes, that is, the main contribution to the entropy of the matter fields, in function of the dimensionless parameter $Mm_{0}$. We can observe the reaching of a steady value. This is a tendency which is being followed by all these large Matsubara modes. Since we have redefined the mass of the black hole as $M = G^{(d)}M_{0}$, we can observe that, for a fixed scalar-field mass, the matter contribution agrees with the generalized second law, and the reaching of a stable value is being driven by the mass of the black hole. The approaching of a constant value evidenced for the entropy contribution from the matter fields could be interpreted as a potential saturation of information on the black hole horizon~\cite{CHEN20151, RAJU20221, 10.21468/SciPostPhys.14.6.150, hartman2013time}. The the stabilization of entropy for large Matsubara modes suggests that high-energy modes contribute less significantly to the overall entropy, which aligns with the ultraviolet cutoff often encountered in different field theories schemes~\cite{10.1143/PTP.14.351, ATHRON2024104094}.

\begin{figure}[ht]
        \begin{center}
            \includegraphics[width=0.7\textwidth]{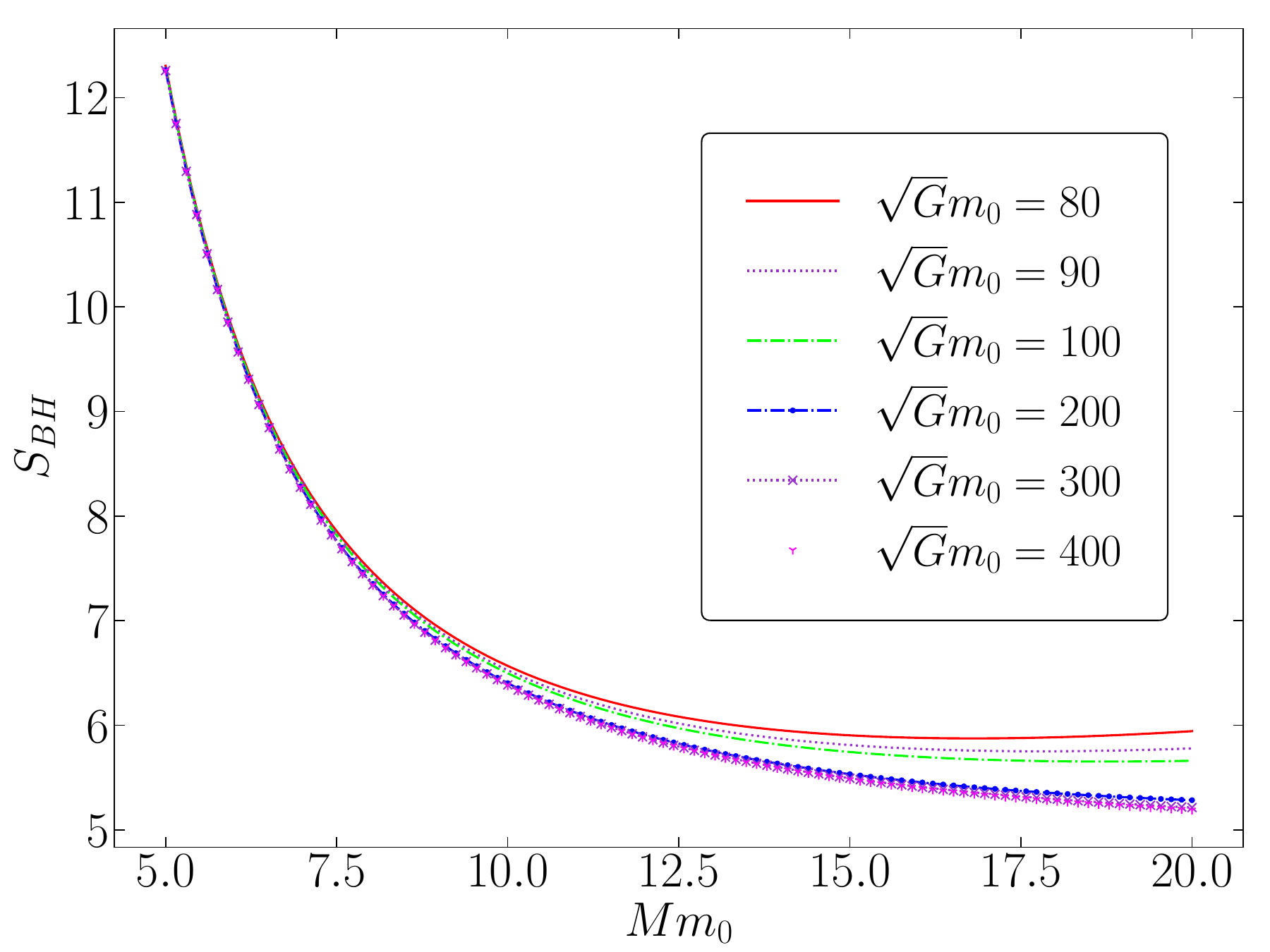}\\
        \end{center}
	\caption{Behaviour of the total black hole entropy $S_{BH}$ as a function of the dimensionless parameter $Mm_{0}$ for different scaled field masses $\sqrt{G}m_{0}$. We do remark the redefinition of $M$ as $M=G^{(d)}M_{0}$.}
	\label{figen2}
 \end{figure}

In Fig. \ref{figen2} we show the validity of the generalized second law of thermodynamics, for different scalar fields, given  by Eq.  (\ref{eq:genentro}), in black hole physics. In other words, we have added the Bekenstein-Hawking entropy to the matter fields entropy described by Eq. (\ref{eq:entro1}) and Eq. (\ref{eq:entro2}), obtaining the expected results. For a range of scalar-field mass values, our findings again confirm the generalized second law, underscoring the robustness of our approach. This demonstrates that the law holds not only for specific cases but across a spectrum of physical parameters. Furthermore, the interaction between the Bekenstein-Hawking entropy and the matter field entropy reveals the intricate balance between geometry and matter in determining the total entropy of the system \cite{bianconi2024gravityentropy, Bianconi_2024}. This balance is essential for understanding the randomness of the degrees of freedom and the thermodynamic properties of black holes within a more comprehensive framework that incorporates both gravitational and quantum effects.

\section{Conclusions}\label{sec:conclusions}
 
The limits of applicability of quantum field theory were put to the test by the formulation of quantum fields in curved spacetime. In this scenario, the concept of black hole entropy was introduced by Bekenstein.
The generalized entropy of the black hole in four dimensions is given by
the area of the event horizon, which is the Bekenstein-Hawking entropy of the system, added to the contribution from the quantized matter and radiation fields.  
To study the generalized second law, we discuss a quantum scalar field in the
Euclidean section of the Schwarzschild manifold, i.e., a quantum field theory analytically continued to imaginary time. We employ this framework to discuss a second contribution, defining the
generalized entropy density of the black hole. 
A conceptually simple way to implement the discussion of Ref. \cite{maldacena} of ``replica wormholes" is to introduce a disorder field linearly coupled with the scalar field. To perform the integration over all the realizations of the disorder, we use the distributional zeta-function method. After integration over all the realizations of the disorder, we obtain a series representation of the averaged generating functional of connected correlation functions, in terms of the moments of the generating functional of correlation functions. Effective actions are defined  
for each of these moments. We show that this approach led us to the theory of Schr\"{o}dinger operators in Riemannian manifolds. 
The necessary and sufficient condition for essential self-adjointness of the generalized Schr\"{o}dinger operator, constructed with the Laplace-Beltrami operator is discussed. 
If it is possible to define self-adjoint operators, the generating functional of connected correlation functions, can be defined. 

Finally, we present the generalized entropy density of the black hole. We evaluated our expressions across various scenarios, showing that the main entropy contribution from the matter and disordered fields reaches a steady value. This stabilization, influenced by the black hole mass, suggests a potential saturation of information interpretation. These findings may align with holographic and wormholes scenarios of interpretation. The addition of Bekenstein-Hawking entropy to the matter field entropy confirms the generalized second law, across a range of scalar-field masses, in the model where the \textit{effects} of the ``replica wormholes" are taken as a quenched disorder over the matter fields.

So far we have considered an additive disorder field. One must consider also a multiplicative disorder situation \cite{Soares_2020}, that can be associated to the coupling between degrees of freedom inside the event horizon with those outside. As has been discussed in the literature, the main difference between the multiplicative and
additive disorder, is that in the 
former, the effects of the disorder fluctuations depend on the state of the system. What comes in mind is if this construction is a particular case of a more general situation discussing quantum fields in bounded domains defined in a Riemannian compact manifold. The literature has been emphasizing that is problematic to define the entropy of quantum field in some compact domain under the influence of an environment. These problems are under investigation by the authors.

\section*{Acknowledgments} 
The authors are grateful to S. A. Dias, B. F. Svaiter, A. M. S. Macedo, C. Farina, G. Krein and R. B. Mann for fruitful discussions. This work was partially supported by Conselho Nacional de Desenvolvimento Cient\'{\i}fico e Tecnol\'{o}gico (CNPq), grant no. 305000/2023-3 (N.F.S.). G.S. thanks to Conselho Nacional de Desenvolvimento Cient\'{\i}fico e Tecnol\'{o}gico (CNPq) due the MSc. scholarship. G.O.H. thanks to Fundação Carlos Chagas Filho de Amparo à Pesquisa do Estado do Rio de Janeiro (FAPERJ) and Coordenação de Aperfeiçoamento de Pessoal de Nivel Superior (CAPES) for the financial support. C.D.R.C acknowledges the funding from the Engineering and Physical Sciences Research Council (EPSRC) (Grants No. EP/R513143/1 and No. EP/T517793/1).

\printbibliography

\end{document}